\documentclass[twocolumn,dvipsnames]{aastex63}

\graphicspath{{./}{figures/}}

\shorttitle{NGC~315}
\shortauthors{Pilawa et al.}

\newcommand{\ngc}{NGC~315}

\newcommand{\kms}{\ensuremath{}{\rm \, km~s^{-1}}}
\newcommand{\msun}{\ensuremath{M_{\odot}}}
\newcommand{\mbh}{\ensuremath{M_\mathrm{BH}}}
\newcommand{\ml}{\ensuremath{M^*/L}}

\newcommand{\GG}[1]{}
\newcommand{\tmaj}{\ensuremath{T_\text{maj}}}
\newcommand{\tmin}{\ensuremath{T_\text{min}}}

\usepackage{xcolor}
\usepackage[flushmargin]{footmisc}

\let\tablenum\relax
\usepackage{siunitx}
\usepackage{amsmath}
\newcommand{\vect}[1]{\boldsymbol{#1}}
\usepackage{graphicx} 
\usepackage{multirow}
\usepackage{booktabs}

\begin{document}

\title{The MASSIVE Survey. XX. A Triaxial Stellar Dynamical Measurement of the Supermassive Black Hole Mass and Intrinsic Galaxy Shape of Giant Radio Galaxy NGC~315}

\correspondingauthor{Jacob Pilawa}
\email{jacobpilawa@berkeley.edu}

\author{Jacob Pilawa}
\affiliation{Department of Astronomy, University of California, Berkeley, CA 94720, USA}

\author{Emily R. Liepold}
\affiliation{Department of Astronomy, University of California, Berkeley, CA 94720, USA}

\author{Chung-Pei Ma}
\affiliation{Department of Astronomy, University of California, Berkeley, CA 94720, USA}
\affiliation{Department of Physics, University of California, Berkeley, CA 94720, USA}

\author{Jonelle L. Walsh}
\affiliation{George P. and Cynthia Woods Mitchell Institute for Fundamental Physics and Astronomy, and Department of Physics and Astronomy, \\
Texas A\&M University, College Station, TX 77843, USA}

\author{Jenny E. Greene}
\affiliation{Department of Astrophysical Sciences, Princeton University, Princeton, NJ 08544, USA}

\begin{abstract}

We present a new dynamical measurement of the supermassive black hole mass and intrinsic shape of the stellar halo of the massive radio galaxy \ngc\ as part of the MASSIVE survey. High signal-to-noise ratio spectra from integral-field spectrographs at the Gemini and McDonald Observatories
provide stellar kinematic measurements in $304$ spatial bins from the central ${\sim}0.3''$ out to $30''$. 
Using ${\sim} 2300$ kinematic constraints, we perform triaxial stellar orbit modeling with the TriOS code and search over ${\sim}$15,000 galaxy models with a Bayesian scheme to simultaneously measure six mass and intrinsic shape parameters. \ngc\ is triaxial and highly prolate, with middle-to-long and short-to-long axis ratios of $p=0.854$ and $q=0.833$ and a triaxiality parameter of $T=0.89$.
The black hole mass inferred from our stellar kinematics is $\mbh = \left(3.0 {\pm} 0.3\right) {\times} 10^{9}\ M_\odot$, which is higher than $\mbh=(1.96^{+0.30}_{-0.13}) {\times} 10^{9} M_\odot$ inferred from CO kinematics (scaled to our distance).
When the seven galaxies with \mbh\ measurements from both stellar and CO kinematics are compared,
we find an intrinsic scatter of 0.28 dex in \mbh\ from the two tracers and do not detect statistically significant biases between the two methods in the current data. The implied black hole shadow size (${\approx} 4.7\, \mu{\rm as}$) and the relatively high millimeter flux of \ngc\ makes this galaxy a prime candidate for future horizon-size imaging studies.

\end{abstract}


\section{Introduction} 

The MASSIVE survey is 
a volume-limited, photometric and spectroscopic survey 
of the ${\sim} 100$ most massive early-type galaxies (with stellar mass $M_* \ga 10^{11.5} \msun$) in the local Universe \citep{maetal2014}. 
A key scientific goal of the survey
is to dynamically measure the masses of a sample of supermassive black holes (SMBHs) within the targeted volume (to a distance of ${\sim}100$ Mpc above declination $\delta = -6^\circ$) with spatially resolved stellar and gas (when present) kinematics. 
To date, 14 MASSIVE galaxies have published dynamical SMBH mass (\mbh) measurements
as compiled in Table 1 of \citet{liepoldma2024}.
Among them, only M87 has \mbh\ determined from the motion of more than one type of dynamical tracers.
For the rest, nine galaxies\footnote{ 
NGC~708, NGC~1453, NGC~1600, NGC~2693, NGC~3842, NGC~4472, NGC~4649, NGC~4889, NGC~7619}
have \mbh\ determined from stellar kinematics using the Schwarzschild orbit modeling method, 
three galaxies\footnote{
NGC~315, NGC~997, NGC~1684}
have \mbh\ inferred from CO kinematics, and one galaxy\footnote{NGC~7052}
is studied with ionized gas. 

In this work, we report a new mass measurement of the SMBH in NGC~315 using stellar kinematics from MASSIVE survey observations and the triaxial orbit modeling method.
NGC~315 has a prior \mbh\ determination based on CO kinematics from Atacama Large Millimeter/submillimeter Array (ALMA)  observations \citep{boizelleetal2021}. NGC~315 is therefore only the second galaxy in the MASSIVE survey for which a direct comparison of \mbh\ from different dynamical tracers can be made.
Beyond MASSIVE, six other galaxies have \mbh\ inferred from both ALMA CO kinematics and stellar kinematics, enabling us to assess the consistency between the two methods (see Sec.~4). \ngc\ is also only the 5th MASSIVE galaxy for which the triaxial stellar orbit modeling is used to determine its \mbh\ (others assumed axisymmetry), and the first MASSIVE galaxy with \mbh\ determined from both CO and {\it triaxial} stellar based methods.

\ngc\ is the brightest member of a galaxy group identified in the Two Micron All Sky Survey (2MASS) group catalog \citep{crooketal2007}. 
The high-density contrast and low-density contrast versions of the catalog list 6 and 97 member galaxies, respectively.
The halo virial mass is estimated to be $3.5{\times} 10^{13} M_\odot$ based on member galaxy velocities.
\ngc\ has strong nuclear radio emission \citep{fanaroffriley1974}  
and a prominent jet extending $\gtrsim100''$ at a position angle (PA) of ${\sim}-50^\circ$
\citep{laingetal2006,riccietal2022}.
\ngc\ is one of seven MASSIVE survey targets with evidence for an X-ray point source in the nuclear region in the $4-7$ keV band. The mean temperature and luminosity of the X-ray hot gas are estimated to be $T_X=0.57$ keV and $L_X=3.8{\times} 10^{41}$ erg s$^{-1}$  \citep{gouldingetal2016}.

Typical of MASSIVE galaxies, \ngc\ is a slow rotator with a velocity amplitude of ${\sim}30 \kms$ and a spin parameter of $\lambda = 0.063$ \citep{vealeetal2017,eneetal2019}.
The PA of the kinematic axis (measured E of N to the receding portion) is determined to be $\text{PA}_\text{kin}= 222^\circ{\pm}7^\circ$ over the $107''{\times} 107''$ field of view (FOV) of the Mitchell integral-field spectrograph (IFS, \citealt{eneetal2018}),
and
$\text{PA}_\text{kin}=218^\circ{\pm} 13^\circ$ in the central $5''{\times} 7''$ region from Gemini IFS data \citep{eneetal2020}.
The ${\sim} 90^\circ$ offset between the kinematic axis and the jet indicates the projection of the angular momentum vector of the stars onto the sky is at the same PA as the jet. 
\textit{HST} photometry shows boxy isophotes and nearly constant ellipticity and photometric PA between a radius of $1''$ and $100''$ with 
luminosity-weighted values of $\epsilon=0.27$ 
and $\text{PA}_\text{phot} = 44.3^\circ{\pm} 0.2^\circ$ (Fig.~1.2 of \citealt{goullaudetal2018}).
The kinematic misalignment angle 
is consistent with 0: $\Psi=6.3^\circ {\pm} 13.3^\circ$ \citep{eneetal2020}.

In Section~2, we discuss the IFS data from Gemini and McDonald Observatories and the stellar velocity moment measurements reported in \citet{vealeetal2017, vealeetal2017a, vealeetal2018, eneetal2018, eneetal2019, eneetal2020}. The \textit{HST} observations of \ngc\ \citep{goullaudetal2018} and surface brightness profile determination \citep{boizelleetal2021} 
are also summarized. In Section~3, we summarize the triaxial orbit modeling code TriOS \citep{quennevilleetal2021, quennevilleetal2022} and the parameter search strategy used to select galaxy models, followed by a discussion of the mass and shape parameters and stellar orbital structure that best match the observations. Section~4 discusses
systematic uncertainties and compares \mbh\ determined from stellar vs. CO kinematics.

Throughout this work, we assume a luminosity distance for \ngc\ of $D_L=68.1{\pm}2.5$ Mpc from the MASSIVE-{\it HST} project using the surface brightness fluctuation technique \citep{goullaudetal2018, jensenetal2021, blakesleeetal2021}. At \ngc's redshift of $z=0.0165$, the corresponding angular diameter distance is $D_A=65.9\pm2.4$ Mpc, and $1''$ is $320$ pc.  
When comparing with \citet{boizelleetal2021}, we adjust their reported values from their assumed $D_A=70$ Mpc to our distance.

\section{Observations}
\label{sec:two}

\subsection{Photometry}

We adopt the characterization of the surface brightness of \ngc\ by \cite{boizelleetal2021} based on archival \textit{HST} and Spitzer photometry.
The \textit{HST} Wide Field Camera 3
(WFC3) observation in the F110W filter cover a $2.1' {\times} 2.2'$ region centered at \ngc\ and produce a final image with $0\farcs08 \text{ pixel}^{-1}$ scale (GO-14219; \citealt{goullaudetal2018}).
Archival Spitzer InfraRed Array Camera 
data from channel 1 ($3.6 \mu m)$ provide deeper coverage of the stellar halo. The final mosaiced image covers a radial range out to $R{\sim}11'$ (about $225$ kpc).

\ngc\ has a prominent circumnuclear dust disk in the central ${\sim}1''$ region.
\cite{boizelleetal2021} 
produce three multi-Gaussian expansion (MGE) models (A, B2, and B3) of the mosaiced image, applying a
differing amount of extinction correction
to the nuclear region.
The central Gaussian component in B2 and B3 has a width of $\sigma^\prime = 0\farcs178$ and $\sigma^\prime = 0\farcs119$, respectively, comparable to the PSF (${\sim}0\farcs15$). As discussed in \cite{liepoldetal2025}, the width of the central component is poorly constrained when it is comparable or below the scale of the PSF. The MGE routine sometimes assigns a central component with $\sigma^\prime \la \sigma^\prime_{\rm PSF}$ that does not improve the fit in a meaningful way but results in an exceptionally large central 3D density after deprojection (recall  $\nu \propto \Sigma / \sigma^\prime$ where $\Sigma$ is the central surface brightnes).
Accordingly, the B2 and B3 MGEs have a ``bump'' in their luminosity densities in the central region, as
shown in Figure~\ref{fig:luminosity_density} in the Appendix.  Additional examples of this artifact can be found in \cite{davidsonetal2024}.
\cite{liepoldetal2025} circumvented this problem by imposing a lower limit on $\sigma'$ during the MGE fitting.
Here, we adopt MGE model A, which has $\sigma^\prime = 0\farcs580$ for the central component and a smooth 3D deprojected density profile without this stellar excess.  In Sec.~4.1 below, we describe results from tests using model B3, the case with the most extreme central luminosity density.

\begin{figure}[htp]
\includegraphics[width=\columnwidth]{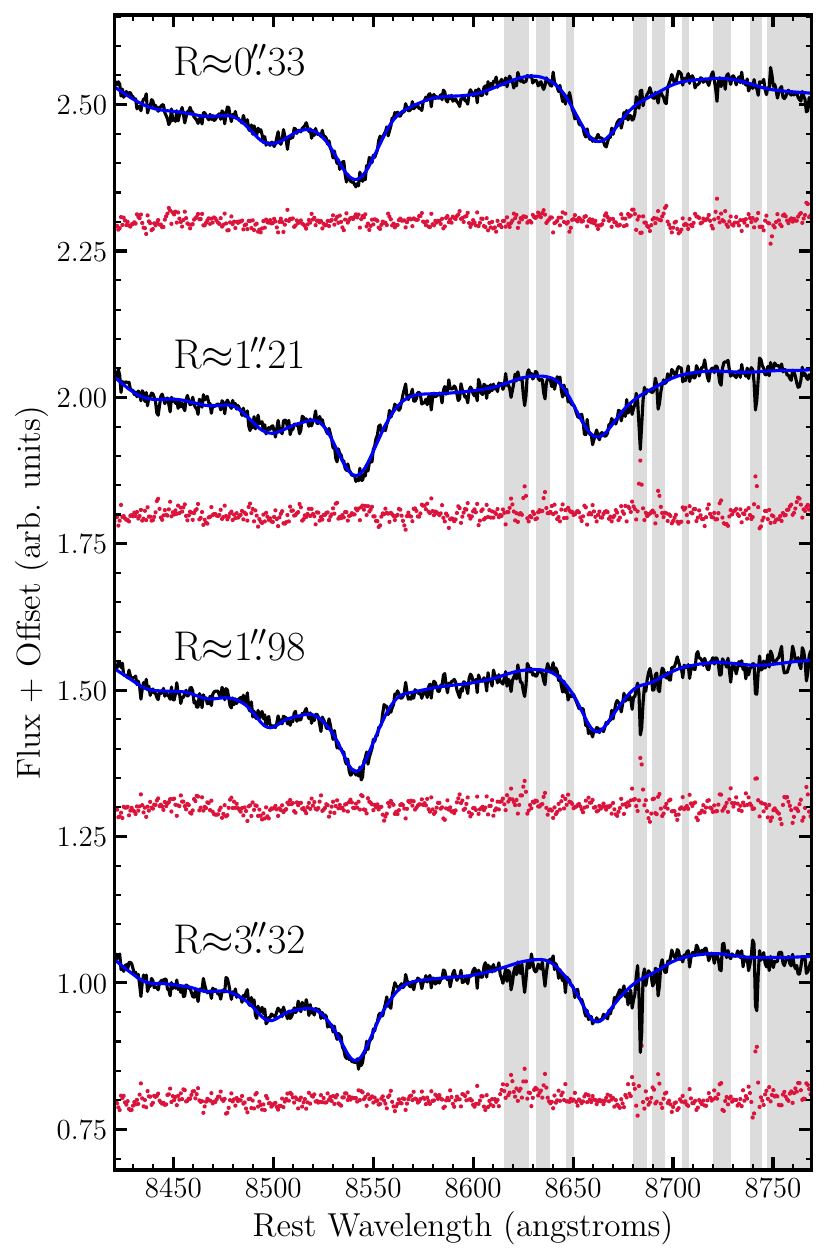}
\caption{
Four representative Gemini GMOS spectra (black) of \ngc\ for spatial bins located at increasing distance from the nucleus. 
The stellar template broadened by the best-fit LOSVD is overlaid (blue) on each spectrum.
The fitting residuals (red points) are offset by constants for clarity. The typical residual is ${\sim}0.5\%$. 
The grey shaded regions are excluded from the fit to account for improperly subtracted sky lines and detector gap. 
}
\label{fig:spectra}
\end{figure}

\subsection{Integral-Field Spectroscopy}

\begin{figure*}[htp]
  \includegraphics[width=6.5in]{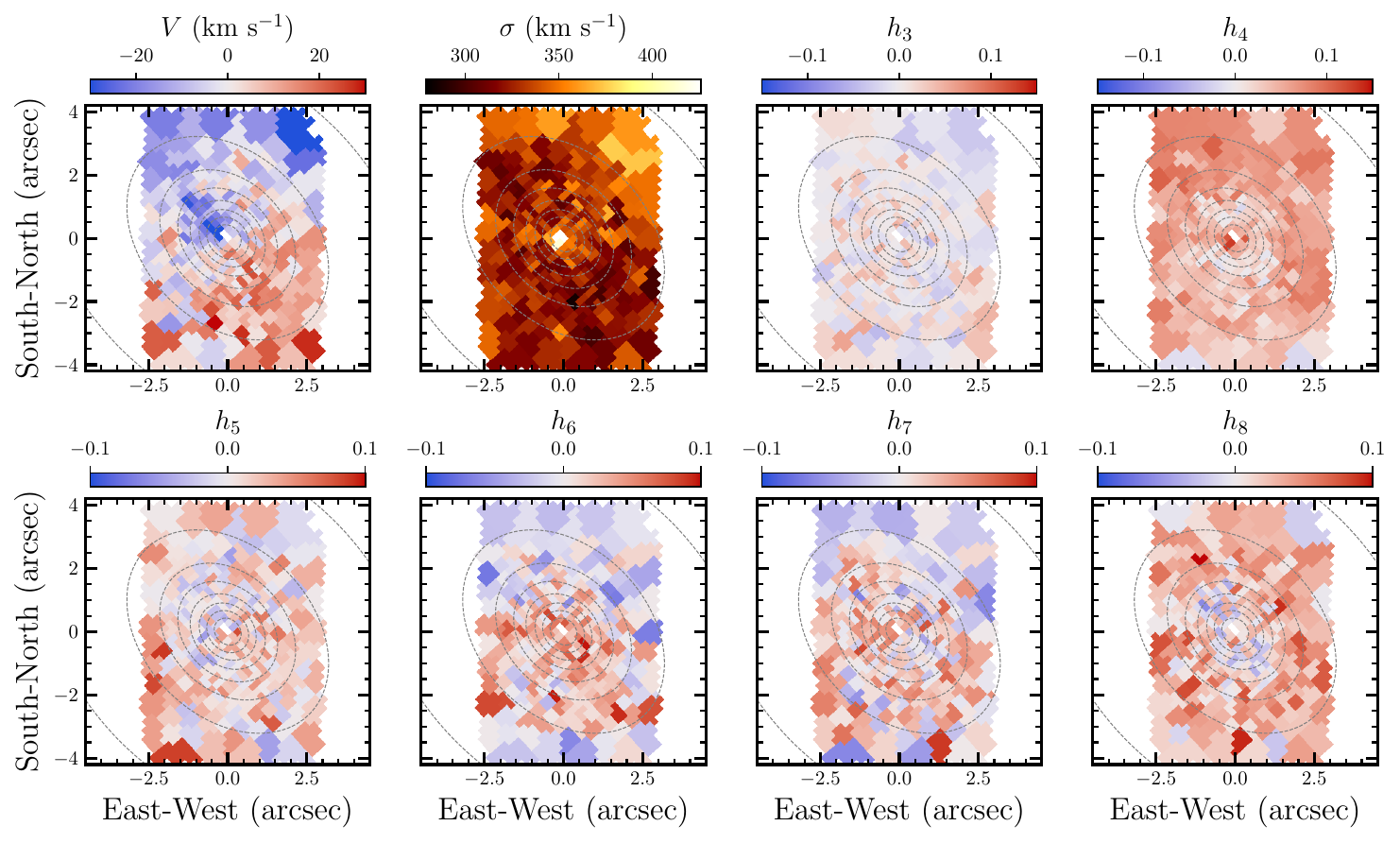}
    \caption{Stellar kinematic maps of the central $5''\times 7''$ region of \ngc\ from Gemini GMOS observations. Spectra from individual lenslets are co-added to achieve a single spectrum with $S/N\ga 125$ for each of the 245 spatial bins.  The two upper-left panels show the line-of-sight velocities $V$ and velocity dispersions $\sigma$, with the higher-order Gauss-Hermite moments $h_3$ to $h_8$ shown in the other panels. 
    Surface brightness contours are plotted as dotted gray lines. 
    }
\label{fig:gmos_map}
\end{figure*}

\subsubsection{Central stellar kinematics} 

We observed the central $5''{\times} 7''$ of \ngc\ using the two-slit IFS mode of Gemini Multi-Object Spectrograph (GMOS) on the Gemini North Telescope with 1000 hexagonal lenslets (each with a projected diameter of $0\farcs2$).
Ten science exposures of 1200 seconds each were obtained, totaling 3.3 hours of on-source and simultaneous observations of a $5'' {\times} 3\farcs5$ region of sky offset by ${\sim}1'$ from the galaxy.
The R$400$-G$5305$ grating with the CaT filter was used to cover the wavelength range $7800{-}9330$~\AA. 
The median spectral resolution (determined from arc lamp lines for each lenslet) was $2.5$ \AA\ FWHM.  
Details of the data processing procedure are described in MASSIVE Paper~XIII \citep{eneetal2019}.

We co-add the spectra from a group of adjacent GMOS lenslets to achieve a threshold signal-to-noise ratio ($S/N$) of 125. This binning procedure
results in 248 high-quality spectra  covering the central region of \ngc.
One difference in this step from \citet{eneetal2019} is that a symmetric binning scheme over the four quadrants of the galaxy was used in that work, while here we perform the binning over the entire GMOS FOV without this assumption. This difference only introduces minor adjustments in how the GMOS lenslets are grouped spatially; the stellar kinematics are measured for each bin without any symmetry assumption in both analyses.

From each spectrum, we measure the line-of-sight velocity distribution (LOSVD) from the CaII triplet (CaT) absorption features over a rest wavelength range of $8420{-}8770$~\AA\ using the penalized pixel-fitting (pPXF) method \citep{Cappellari2017}. 
The LOSVD is represented as a Gauss-Hermite series of order $n=8$. 
We use $15$ stellar templates 
from the MILES CaT Library  that covers a wavelength range of $8437{-}9020$~\AA\ with a spectral resolution of $1.5$~\AA\ FWHM \citep{Cenarroetal2001}. 
This set of stars is taken from Table 2 of \cite{Barthetal2002}, but
we find consistent stellar kinematics when the full library of $706$ stars
is used. 
A multiplicative polynomial of degree three is used to model the stellar continuum in each spectrum.

\begin{figure*}[ht]
  \centering
  \includegraphics[width=0.95\linewidth]{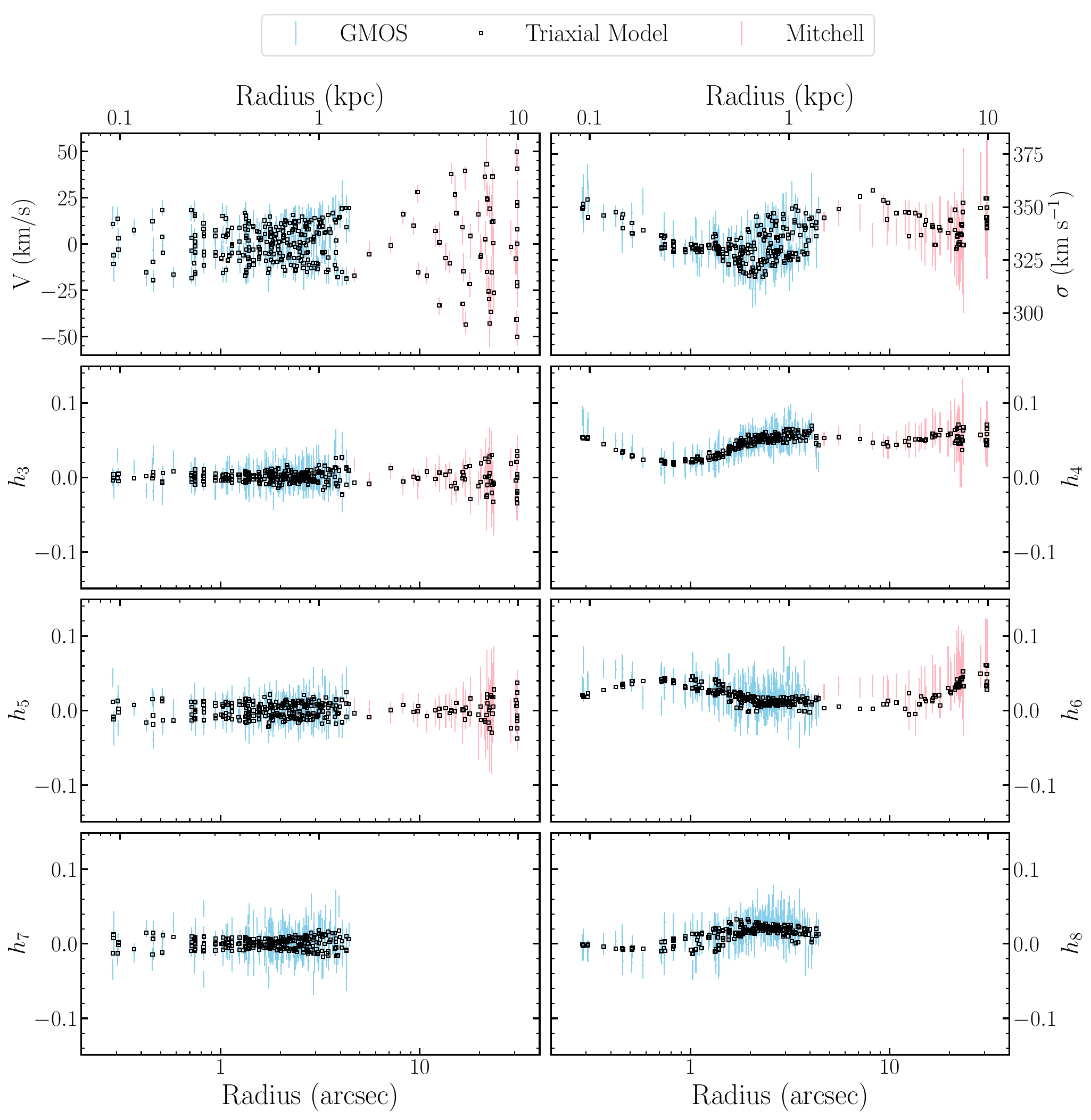}
  \hfill
  \caption{
  Radial profiles of the stellar kinematic moments for \ngc\ from Gemini GMOS (blue bars) and Mitchell (pink bars) data. Moments predicted by the best-fitting triaxial galaxy model (black squares) listed in Table~1 match the input data well. 
  }
\label{figure:moments}
\end{figure*}

Four representative spectra at increasing radii are plotted in Figure~\ref{fig:spectra}. 
The template spectrum broadened by the best-fit LOSVD (blue curve) provides an excellent fit to each observed spectrum (black curve) with a typical residual (red points) of ${\sim} 0.5$\%.
In addition to the bright sky lines, the wavelength range ${\gtrsim}8750$~\AA\ is also masked due to the presence of a gap in the CCD chip. The resulting maps of the eight Gauss-Hermite moments of the LOSVDs for the 248 GMOS spatial bins are shown in Figure~\ref{fig:gmos_map}. The corresponding radial profiles of the eight moments are displayed as blue bars in Figure~\ref{figure:moments}.
The errors on the moments are determined via the Monte Carlo method described in Section~4 of \cite{eneetal2019}.
These figures show a velocity profile with a low amplitude rotational velocity $|V| {\sim} 20 \,\kms$, and a velocity dispersion profile that rises from $\sigma {\sim} 320 \kms$ at $R {\sim} 2''$ inward to $\sigma {\sim} 350 \kms$ at $R {\sim} 0\farcs3$.

\subsubsection{Wide-field stellar kinematics}

We observed \ngc\ with the Mitchell IFS at the McDonald Observatory as part of the MASSIVE Survey. Details of the observation, data reduction and kinematic measurements are described in \cite{vealeetal2017, vealeetal2017a, vealeetal2018}.
The observations consisted of three dither positions, during which we interleaved two $20$ minute science frames with one $10$ minute sky frame, resulting in $2$ hours on-source. Each frame spans a $107''{\times} 107''$ FOV with $246$ fibers, covering a wavelength range of $3650{-}5850$~\AA\ that includes the Ca HK region, the $G$-band region, $H\beta$, Mg$b$, and several Fe lines. The individual fibers in the central region of \ngc\ provide spectra with $S/N \gtrsim 50$. For the fainter part of the galaxy covered by the outer fibers, the spectra are co-added to achieve $S/N \gtrsim 20$. 

The LOSVD from each Mitchell spectrum is extracted in a similar way as GMOS described above.  We opt to fit to $n=6$ Gauss-Hermite moments due to the lower $S/N$ data here. The MILES library of $985$ stellar spectra is used as templates \citep{Sanchez-Blazquezetal2006, Falcon-Barrosoetal2011}. The kinematic moments for the 55 Mitchell bins are shown in Figure~\ref{figure:moments} in pink. The Mitchell data points connect smoothly to the GMOS points, showing excellent agreement between measurements obtained from different spectrographs, telescopes, and spectral regions.  Six additional kinematic points at $R{\sim} 50''$ are shown in Figure~D1 of \citet{vealeetal2017} but are excluded in the following analysis due to the low $S/N$ of these outer spectra. 

\section{Results from Triaxial Orbit Modeling}
\label{sec:three}

\subsection{The TriOS Code and Galaxy Models}

We use the \textit{TriOS} code \citep{vandenboschetal2008,quennevilleetal2021,quennevilleetal2022} to compute triaxial orbit models of \ngc. This code integrates a large number of stellar orbits that span the allowed phase space and computes the LOSVDs for a wide range of galaxy model parameters.
The galaxy is assumed to have three mass components: a central black hole of mass $\mbh$, a stellar component with a mass-to-light ratio $\ml$, and a dark matter halo with a density profile 
$\rho(r) =\rho_0/[(r/r_s)^\gamma\left(1+r/r_s\right)^{3-\gamma}]$,
where 
$r_s$ is a scale radius \citep{navarroetal1996}. We set $\gamma = 0$ so that the profile has a finite central density $\rho_0$ and a flattened central 
density distribution similar to that of the stars.
With our data, $r_s$ and $\rho_0$ are often quite degenerate
so we choose to parameterize the halo with a single parameter, $M_{15}$, defined to be the dark matter mass enclosed within $15$ kpc with a fixed scale radius $r_s = 15$ kpc. A similar strategy was used in orbit modeling of other MASSIVE galaxies (e.g., \citealt{liepoldetal2020, quennevilleetal2022, pilawaetal2022}).

We use three parameters to specify the triaxial shape of the stellar component: $p= b/a$ is the intrinsic middle-to-long axis ratio, $q = c/a$ is the intrinsic short-to-long axis ratio, and $u$ is the apparent-to-intrinsic long axis ratio. These three shape parameters are related to the three angles $\theta$, $\phi$, and $\psi$ that relate the intrinsic and projected coordinate systems of \ngc; see Equations~(4) and (8) of \citet{quennevilleetal2022}. Here $\theta$ and $\phi$ are polar angles in \ngc's intrinsic coordinate system, and $\psi$ specifies the remaining degree of freedom -- a rotation of the galaxy around the line of sight.

In total, each galaxy model has $6$ free parameters -- $\mbh$, $\ml$, $M_{15}$, and three shape parameters -- to be constrained by the kinematic and photometric data. 
For each model, 
we use the same procedures for phase space sampling and orbit integrations as in our earlier work (e.g., Section~4.1 of \citealt{liepoldetal2023}).
We integrate the trajectories of stars to build a library of 437,400 stellar orbits. A dithering factor of $N_{\rm dither}=3$ is used, where the properties of $N_{\rm dither}^3 = 27$ orbits with similar initial conditions are averaged, effectively giving them a single shared orbital weight.
We determine the resulting 16,200 independent orbital weights using the kinematic measurements with non-negative least squares \citep{lawsonhanson1995}, under the additional constraint that both the projected mass within each aperture and the 3D mass distribution in coarse bins are consistent within 1\% of the MGE.
The GMOS and Mitchell PSFs are taken to be single, circularly symmetric Gaussians with FWHM of $0\farcs38$ and $1\farcs2$, respectively.

\begin{figure}[ht]
  \centering
  \includegraphics[width=3.8in]{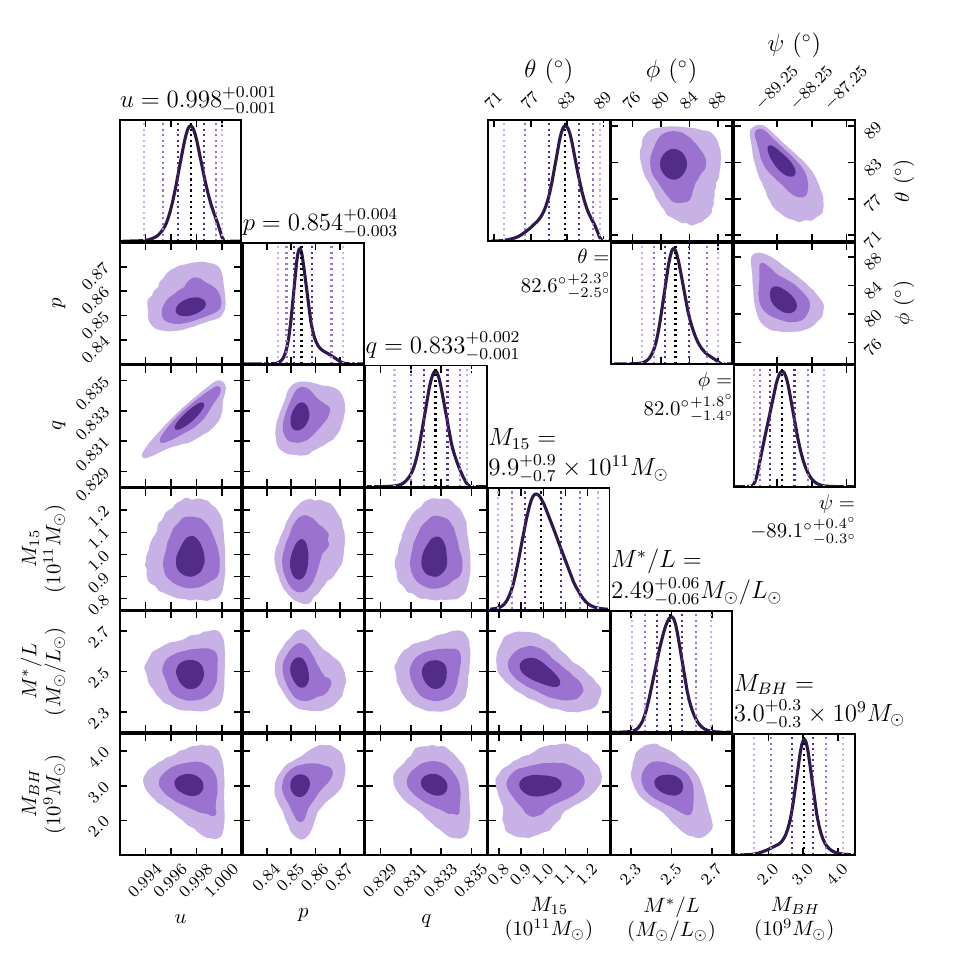}
  \caption{(Left) Posterior distributions of six parameters in triaxial orbit modeling of \ngc: SMBH mass \mbh, stellar mass-to-light ratio \ml, dark matter enclosed within 15 kpc $M_{15}$, and luminosity averaged axis ratios $u, p$, and $q$. The 68\%, 95\%, and 99.7\% credible 
  regions are represented by different shades of purple.
  The vertical lines in each 1D marginalized distribution indicate the median and the three corresponding confidence levels. (Upper right) Posterior distributions in viewing angle space, where 
  $\theta$ and $\phi$ are polar angles in the galaxy's frame, and $\psi$ specifies a rotation of the galaxy around the line of sight (Section~2.2 of \citealt{quennevilleetal2022}).
  } \label{fig:CornerPlots}
\end{figure}

\begin{table}[ht]
\hspace{-3em}
\begin{tabular}{>{\scriptsize}l >{\scriptsize}r}
\multicolumn{2}{c}{{\bf Table~\ref{tab:triax_results}}}\\[0.5ex]
\multicolumn{2}{c}{Galaxy Parameters of \ngc\ from Triaxial Orbit Modeling}\\[1ex]
\hline\hline
Galaxy Property (units)                    & Inferred Value                     \\ \hline
Black hole mass $\mbh$ $(10^{9} M_\odot)$                   & $3.0{\pm}0.3$      \\
$\ml$ ($M_\odot / L_\odot$)                   & $2.49{\pm}0.06$    \\
Total stellar mass ($10^{12} M_\odot$)                            & $1.5$    \\
DM mass within 15 kpc $M_{15}$ $(10^{11} M_\odot)$                  & $9.9^{+0.9}_{-0.7}$    \\
Triaxiality parameter $T$                                          &   
$0.89{\pm} 0.02$\\
Shape parameter $\tmaj$                                          &  $0.019^{+0.008}_{-0.007}$             \\
Shape parameter $\tmin$                                          & $0.017^{+0.009}_{-0.013}$              \\
Average middle-to-long axis ratio $p$                                          & $0.854^{+0.004}_{-0.003}$            \\
Average short-to-long axis ratio $q$                                          & $0.833^{+0.002}_{-0.001}$            \\
Average apparent-to-intrinsic long axis ratio $u$                                          & $0.998{\pm}0.001$            \\
Line-of-sight direction $\theta (^\circ)$, $\phi (^\circ)$                & $82.6^{+2.3}_{-2.5}$, $82.0^{+1.8}_{-1.4}$                       \\
Rotation about line of sight $\psi \left(^\circ\right)$                   & $-89.1^{+0.4}_{-0.3}$                        \\ \hline
\end{tabular}
\caption{
For each parameter, we marginalize over the other parameters and report the
68\% credible regions.
In orbit models, $\theta$ is the inclination angle in the oblate axisymmetric limit ($\psi=90^\circ$, or equivalently $p=1$), with $\theta=90^\circ$ being edge-on and $\theta=0^\circ$ being face-on.}
\label{tab:triax_results}
\end{table}

\subsection{Best-Fit Triaxial Model}

We use the model selection scheme described in our recent work (e.g., \citealt{pilawaetal2022, liepoldetal2023, liepoldetal2025})
to determine the mass and shape parameters that best match the kinematic and photometric data of \ngc. This scheme, in brief, involves generating candidate galaxy models with Latin hypercube sampling \citep{mckayetal1979}, approximating the resulting likelihood landscapes with Gaussian process regression, and sampling posteriors on our parameters via dynamic nested sampling \citep{speagle2020}. The relative likelihood for each model is computed from
\begin{equation}\label{eqn:logl}
-2 \ln \mathcal{L}(\vect{\mu}) = \chi^2_\textrm{kin}(\vect{\mu}) = \sum_j^{N_\textrm{bin}}\sum_i^{N_\textrm{GH}} \frac{(h_{ij}(\vect{\mu}) - h_{ij,\rm{data}})^2}{\Delta h_{ij,\rm{data}}^2} \,,
\end{equation}
where $h_{ij}$ is the $i$-th Gauss-Hermite moment of the stellar
LOSVDs in the $j$-th spatial bin, $\Delta h_{ij}$ is the measurement uncertainty, and $\vect{\mu}$ is the set of six parameters describing the galaxy's potential. 
To obtain 1D confidence intervals for each galaxy parameter, we take the marginalized distribution for each parameter to be the distribution of (weighted) sample points over the desired parameters, assuming uncertainties in the distribution function (e.g., orbital weights) are neglected.

About ${\sim} 1750$ galaxy models are used to obtain the final posteriors
on the six parameters shown in Figure~\ref{fig:CornerPlots}.
An additional ${\sim} 3000$ models covering wider ranges of parameters are used in the initial exploration of the 6D likelihood surface, and ${\sim} 10000$ more models are used in various tests to ensure that the surface in the low $\chi^2$ region is mapped out accurately. 

A summary of the best-fit parameters for \ngc\ is listed in Table~\ref{tab:triax_results}. The kinematic moments predicted by the model are compared with observed values in Figure~\ref{figure:moments}. 
The total $\chi^2$ is $1611.4$ spread over $2314$ kinematic constraints ($8$ moments for 248 GMOS bins and $6$ moments for 55 Mitchell bins).
A naive estimate of the 
reduced $\chi^2$ would be $1611.4/(2314-6)=0.698$ assuming 6 degrees of freedom (DOF). 
But we caution that DOF is nontrivial
to estimate for nonlinear problems such as here.  In a study using simulated stellar kinematics that mimic data in the MASSIVE survey \citep{pilawaetal2024}, we have investigated a ``generalized" measure of DOF in the form of a penalty term added to the likelihood measure \citep{ye1998, lipkathomas2021}
and found it to be ${\sim 200}$ instead of the canonical value of 6, thereby raising the reduced $\chi^2$ by ${\sim} 40$\% in that study.
While a similar calculation would have to be performed to estimate this alternative DOF measure for \ngc, we expect the reduced $\chi^2$ to be raised as well.  
Using this generalized measure to estimate the effective DOF in the models also helps address a related concern that the conventional (unpenalized) likelihood method does not marginalize over the high dimensional space of possible  
orbital weights, leading to reliable \mbh\ but with ``overly pessimistic error estimates" \citep{magorrian2006}.
In \citet{pilawaetal2024}, we find that the parameter search and inference routines used in this paper and other work by our group (e.g., \citealt{quennevilleetal2022, pilawaetal2022, liepoldetal2023, 
pilawaetal2024, liepoldetal2025})
are able to reliably recover the mass and shape parameters associated with synthetic galaxy kinematic data without any additional penalty terms.

A handful model predictions in Figure~\ref{figure:moments} lie outside the 68\% uncertainties of the data points, e.g.,
the $h_{4}$ and $h_{6}$ GMOS moments within $0\farcs6$ are under-predicted (on average) by ${\sim}1.05\sigma$ and ${\sim}1.11\sigma$, respectively, and Mitchell's velocity dispersion and $h_{6}$ between $5''$ and $10''$ are over- and under-predicted by ${\sim}1.1\sigma$ and ${\sim}1.6\sigma$, respectively.
These systematic local deviations occur in $\lesssim 2\%$ of the 2314 kinematic constraints. When excluding these deviant constraints from the $\chi^2$ used for our parameter inference, we find an insignificant change in the inferred \mbh.

\subsubsection{Black Hole Mass, Stellar Mass, and Dark Matter Mass}

To assess how \ngc\ and its SMBH fit in the population of local galaxies with dynamically inferred \mbh, we examine its location on the well-studied $\mbh{-}\sigma$ and $\mbh{-}M_{\rm bulge}$ relations.  
\ngc's luminosity-weighted velocity dispersion within $R_e$ is found to be $\sigma_e=341 \text{ km s}^{-1}$ based on the same Mitchell IFS data in this paper \citep{vealeetal2017}.
At this $\sigma_e$, the mean full-sample $\mbh{-}\sigma$ relations of \citet{mcconnellma2013} and \cite{Sagliaetal2016} predict
$\mbh=4.2\times 10^{9} M_\odot$ and $3.3\times 10^{9} M_\odot$, respectively,
$40\%$ (0.15 dex) and $10\%$ (0.04 dex) larger than our dynamically measured \mbh. However, our \mbh\ is within the intrinsic scatter of both relations, $0.38$ dex.

To place \ngc\ on the $\mbh{-}M_{\rm bulge}$ relation, we use the total stellar mass from our best-fitting triaxial model, $M_{*} = 1.5\times 10^{12} M_\odot$. At this bulge mass,
the mean full-sample relations of  \cite{mcconnellma2013} and \cite{Sagliaetal2016} predict $\mbh=5.3\times10^{9}M_\odot$ and $4.4\times10^{9}M_\odot$, respectively, $77$\% (0.25 dex) and $47$\% (0.17 dex) larger than our measured \mbh.  But again, our dynamical \mbh\ is within the intrinsic scatter of both $\mbh{-}M_{\rm bulge}$ relations, $0.34$ dex.

In our preferred model, the black hole has a gravitational sphere of influence (SOI) of $r_\text{SOI}=0\farcs81 = 0.26$~kpc with the definition of $M^{*}(<r_\text{SOI}) = \mbh$, and $r_\text{SOI}=1\farcs1 = 0.35$~kpc with the definition of $M^{*}(<r_\text{SOI}) = 2\mbh$.

Within the effective radius $R_e=25\farcs7$ (${\sim}8.3$ kpc) of \ngc\ \citep{quennevilleetal2024}, the dark matter halo constitutes ${\sim}47\%$ of the total mass of the galaxy. 
This value is broadly consistent with the rising dark matter fraction with increasing galaxy stellar mass reported for lower-mass early-type galaxies (e.g., Fig.~10 of \citealt{cappellarietal2013}; Fig.~16 of \citealt{santuccietal2022}), where the dark matter fraction (at $1\,R_e$) reaches ${\sim}40\%$ at the highest masses ($M^\star{\sim } 5\times 10^{11}M_\odot$) in their samples.

\subsubsection{Intrinsic 3D Galaxy Shape}

The preferred axis ratios of $p=0.854$ and $q=0.833$ 
yield a triaxiality parameter of $T = \left(1-p^2\right)/\left(1-q^2\right)= 0.89{\pm} 0.02$, where the limits of $T=0$ and 1 correspond to oblate axisymmetry and prolate axisymmetry, respectively. 
In comparison, five other massive elliptical galaxies for which we have performed triaxial orbit modeling thus far all have smaller $T$. Four of them are oblate ($T \la 0.5$): $T = 0.33{\pm} 0.06$ for NGC~1453 \citep{quennevilleetal2021}, $T=0.39{\pm} 0.04$ for NGC~2693 \citep{pilawaetal2022}, $T=0.35{\pm} 0.03$ for Holmberg~15A \citep{liepoldetal2025}, and  $T=0.31{\pm} 0.05$ for NGC~57 (Pilawa et al.~2025, in prep).  
M87, on the other hand, is slightly prolate with $T=0.65{\pm} 0.02$ \citep{liepoldetal2023}.

While the determinations of the intrinsic axis ratios $p$ and $q$ for individual galaxies require triaxial orbit modeling and detailed kinematic data, one can infer the distributions of $p$ and $q$ {\it statistically} from the observed isophotal shapes and kinematic vs. photometric misalignment angles of an ensemble of galaxies. For slow-rotating galaxies in the MASSIVE survey, the mean values are found to be $\langle p\rangle = 0.88$, $\langle q\rangle=0.65$, and $\langle T \rangle = 0.39$ \citep{eneetal2018}, similar to the majority of our directly measured $T$ thus far.
Comparable distributions are also found for early-type galaxies in the ATLAS$^{\rm 3D}$ and SAMI surveys \citep{Weijmansetal2014,Fosteretal2017} and for simulated massive slow rotators in the IllustrisTNG simulations \citep{pulsonietal2020}.
We therefore expect stellar halos with the prolateness of \ngc\ to be a rare occurrence.

\subsubsection{Stellar Orbital Structure}

\begin{figure}[htp]
\includegraphics[width=\columnwidth]{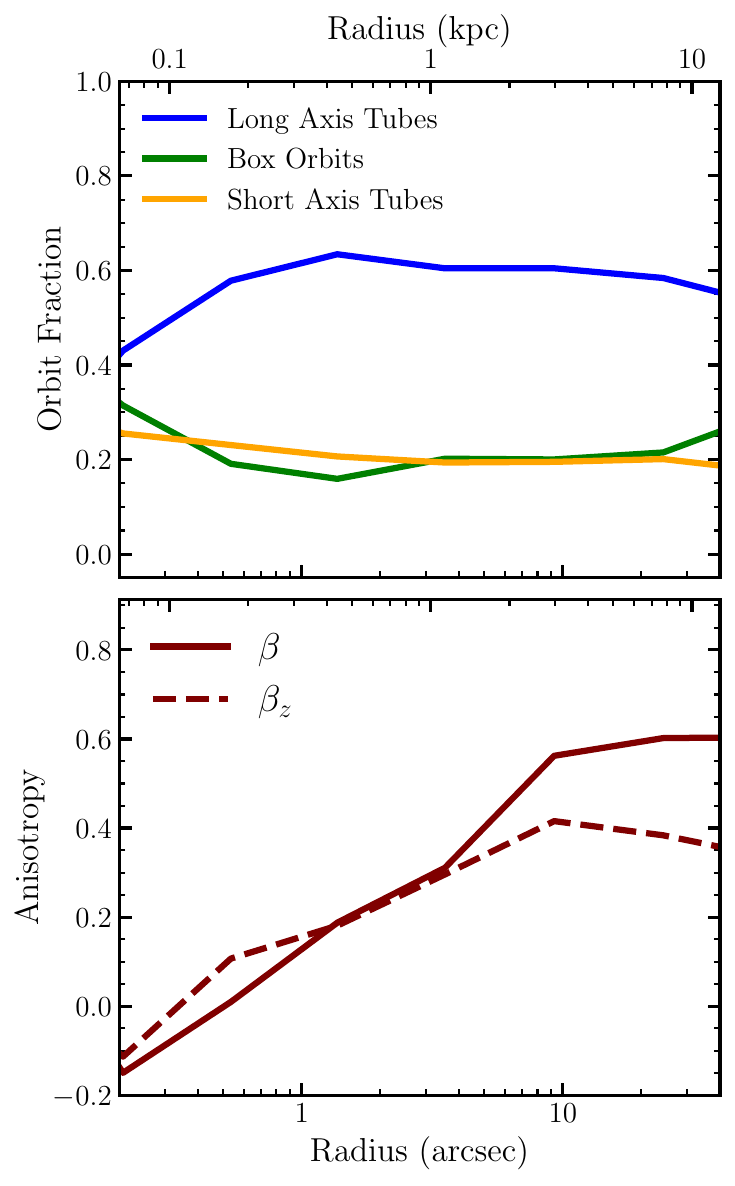}
\caption{ 
Composition
of the three major orbit types (upper panel) and velocity anisotropy (lower panel) as a function of radius in the best-fit triaxial galaxy model of \ngc. The majority of the orbital weights are in long-axis tube orbits, consistent with the prolateness of the galaxy. Short-axis tube orbits and box orbits both contribute about $20\%-25$\% to the remaining orbital weights.}
The velocity anisotropy parameters
$\beta \equiv 1 - \sigma_t^2 / \sigma_r^2$ (in spherical coordinates) and $\beta_z = 1 - \sigma^2_z/\sigma^2_R$ (in cylindrical coordinates) 
indicate the stellar orbits are mildly tangentially anisotropic at small radii and are increasingly radially anisotropic at larger radii. 
\label{fig:orbital_comp}
\end{figure}

The composition of the major types of orbits in the best-fit model for \ngc\ is plotted as a function of radius in the upper panel of Figure~\ref{fig:orbital_comp}. The two types of tube orbits have a fixed sense of rotation, with short- and long-axis tubes having angular momentum components along the intrinsic short- and long-axis, respectively, 
which do not change sign.
For the box orbits, all three components of the angular momentum change sign, leaving no sense of rotation for this orbit type. The relative contribution of these three orbit types determines the velocity structure of the galaxy.

Figure~\ref{fig:orbital_comp} shows that $45\%-60\%$ of the mass in the orbits are from
long-axis tubes, while short-axis tubes and box orbits contribute comparable amounts to the remaining orbital weight. 
The predominance of long-axis tubes over short-axis tubes reflects the prolateness of \ngc.
This is opposite to the orbital compositions in oblate triaxial galaxies with $T<0.5$, e.g., about $60{-}70$\% of orbits are short-axis tubes while ${\sim} 20$\% are long-axis tubes in NGC~1453 and NGC~2693 \citep{quennevilleetal2021, pilawaetal2022}. 

Further insight into the relative short- and long-axis tube orbits can be gained from the relationship between the parameter $T$ and the locations in the start space from which the tube orbits are generated (using the scheme of \citealt{schwarzschild1993}). \citet{quennevilleetal2021} shows that the location of the focal curve demarcating the regions in start space populated by long- and short- axis tubes (their Fig.~1) is specified by an angle $\eta$, where 
$\tan \eta \approx \sqrt{T/(1-T)}$.  For $T=0.89$, we have $\eta\approx 71^\circ$, indicating the long- and short-axis tube fraction in the start space is 
$79$\% and $21$\%, respectively, corresponding to a ratio of about 4 to 1.
The actual orbital fractions for the best-fit galaxy model shown in Figure~\ref{fig:orbital_comp} are computed from these base orbits with the corresponding orbital weights, yielding a comparable long-to-short tube ratio of roughly $3$ to $1$.\footnote{The preferred viewing angles (in Table~\ref{tab:triax_results}) are close to the intermediate axis ($\theta$ = $\phi$ = $90^\circ$). In this limit, the freedom for the deprojection extends roughly along the intermediate axis, corresponding to different values of $T$ (see Equations~(7) and (8) of \citealt{quennevilleetal2022}). Within the viewing-angle parameterization of the deprojection, the third viewing angle $\psi$ then sets $T$.}

The lower panel of Figure~\ref{fig:orbital_comp} displays the radial profile of the velocity 
anisotropy parameters, $\beta = 1-\sigma_t^2/\sigma_r^2$ and $\beta_z = 1-\sigma_z^2/\sigma_R^2$, where $\sigma_t$ and $\sigma_r$ are the tangential and radial velocity dispersions in spherical coordinates, and $\sigma_z$ and $\sigma_R$ are the vertical and radial velocity dispersions in cylindrical coordinates, respectively.
The orbits in the central portion of \ngc\ are slightly tangential with $\beta<0$ and become radially anisotropic away from the center.
This trend is typical in massive elliptical galaxies (e.g., \citealt{Thomasetal2016}) and can be seen in other MASSIVE galaxies (e.g., \citealt{liepoldetal2020, pilawaetal2022}).

\section{Discussion}

\subsection{Uncertainties in Central Surface Brightness}

A primary systematic uncertainty in the mass measurements of \ngc\ SMBH is the effect of dust on the observed central stellar light. 
In the main analysis above, we have used MGE model A of \citet{boizelleetal2021} to approximate the surface brightness profile of \ngc.
To test the impact of the adopted profile on the inferred \mbh,
we replace it with their MGE B3, the model that assumes the largest extinction correction.
We re-run orbit models and parameter search in the reduced 3D space of \mbh, \ml, and halo mass with the shape parameters fixed to the best-fitting values. While model B3 has a significantly higher central luminosity density than model A (Figure~\ref{fig:luminosity_density}), the difference is confined to the inner ${\sim}0\farcs5$, a scale comparable to the PSF of our Gemini observations.
It is thus not surprising that our
tests find ${\lesssim 3}$\% changes in the best-fitting parameters, with  $\mbh = (2.9\pm 0.2)\times 10^9 M_\odot$ and $\ml = 2.45\pm0.04\, M_\odot/L_\odot$.
In comparison, \citet{boizelleetal2021} find that switching from model A to B3 in their CO-based study reduces \mbh\ and \ml\ by ${\sim}18\%$. One possible reason for this larger change in their tests is the CO kinematic data only extend to ${\sim} 1''$,
and thus their constraints are more sensitive to changes in the central luminosity density and the enclosed mass.

\begin{figure}[htp]
  \includegraphics[width=\columnwidth]{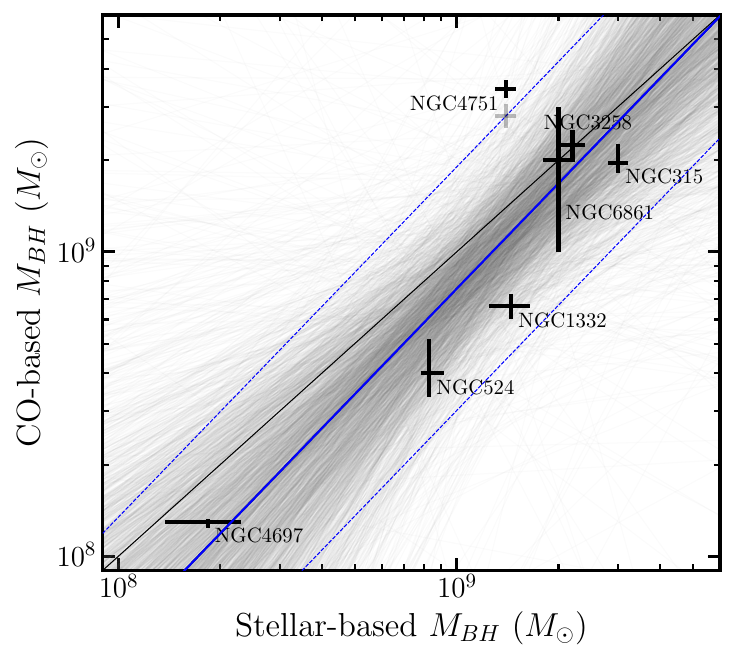}
    \caption{
    Comparison of seven local galaxies with \mbh\ determined independently from stellar and CO kinematics listed in Table~2. 
   The median linear fit (solid blue line) is given by $\log_{10} (\mbh^{\rm CO} /10^9 M_\odot ) = (1.15^{+0.44}_{-0.41}) \log_{10} (\mbh^{\rm stellar}/10^9 M_\odot) - (0.12 {\pm} 0.15)$, with an intrinsic scatter of $\epsilon = 0.35^{+0.26}_{-0.12} \text{ dex}$ 
   (dotted blue lines). Faint gray lines are constructed from sampling over the posteriors for our parameters.
    The solid black line denotes the one-to-one line to guide the eye. 
    The two points for NGC~4751 represent $\mbh$ inferred when assuming a spatially varying (lower point) vs. constant (upper point) $\ml$ in the CO-based model (see text).}
    \label{fig:comparison} 
\end{figure}

\subsection{\mbh\ from Stellar vs. CO Kinematics}

With the mass measurement of \ngc's SMBH presented in this paper, there are now seven galaxies whose SMBH masses have been determined from the kinematics of both CO and stars (with orbit-based modeling).
A summary of these measurements is given in Table~\ref{tab:summary} and Figure~\ref{fig:comparison}. 
To assess the level of agreement between \mbh\ from the two methods, we perform a linear fit using the \texttt{LinMix} package \citep{kelly2007}. This method derives the best-fitting line through hierarchical Bayesian modeling, accounting for uncertainties in both $x$ and $y$ directions. It models the distribution of the independent variable as a mixture of Gaussians and assumes the dependent variable is drawn from a Gaussian distribution centered on a linear relation with the independent variable. The code returns posterior distributions for the parameters, including an intrinsic scatter term representing the variance beyond measurement uncertainties.

The resulting linear fit for \mbh\ from the two methods is $\log_{10} (\mbh^{\rm CO} /10^9 M_\odot ) = (1.15^{+0.44}_{-0.41}) \log_{10} (\mbh^{\rm stellar}/10^9 M_\odot) - (0.12{\pm} 0.15)$, 
with an intrinsic scatter of $\epsilon = 0.35^{+0.26}_{-0.12}$ dex (blue lines in Figure~\ref{fig:comparison}).  
To assess whether the data prefer a one-to-one model with a fixed slope of $1$ and intercept of $0$ or a linear relation with free coefficients, we compare the marginal evidence ratio (``Bayes factor'') of the two models.
The one-to-one model here is a subset of the free linear model, so the Bayes factor in this case quantifies the factor by which a free slope and intercept improves the fit. Using \texttt{dynesty}, we find a log-marginal evidence difference of $\Delta \ln Z = 2.78 \pm 0.02$ (one-to-one model minus free model), which is much less than what is typically considered evidence in favor of one model (e.g., $|\Delta \ln Z| > 5$, \citealt{trotta2008,lockhartgralla2022}). 
We therefore conclude that 
while there are differences in individual \mbh\ values determined from CO and stellar kinematics, there is no evidence for statistically significant biases between the two methods in the current data.

{
\setlength\LTcapwidth{\textwidth} 
\begin{longtable*}[c]{lcccccccl}
\multicolumn{9}{c}{{\bf Table~\ref{tab:summary}}}\\
\multicolumn{9}{c}{Galaxies with dynamical \mbh\ measurements from both stellar and CO kinematics}\\
\hline\hline
\multicolumn{1}{c}{} & \multicolumn{4}{c}{Stellar-Based Measurements} & \multicolumn{3}{c}{CO-Based Measurements} & \multicolumn{1}{c}{} \\
\cmidrule(l{0.5em}r{0.5em}){3-5} \cmidrule(l{0.5em}r{0.5em}){6-8}
Name & $D$ & $M_{\rm BH}$ & $M^{*}/L$ & $i$ & $M_{\rm BH}$ & $M^{*}/L$ & $i$ & Ref. \\
     & (Mpc) & ($10^{9}\,M_\odot$) & [band] &  & ($10^{9}\,M_\odot$) & [band] &  &  \\
(1)  & (2)   & (3)                  & (4) & (5) & (6)                & (7) & (8) & (9) \\ \hline
\endfirsthead

\hline\hline
Name & $D$ & $M_{\rm BH}$ (stellar) & $M^{*}/L$ [band] & $i$ & $M_{\rm BH}$ (CO) & $M^{*}/L$ [band] & $i$ & Ref. \\
     & (Mpc) & ($10^{9}\,M_\odot$) &  &  & ($10^{9}\,M_\odot$) &  &  &  \\
(1)  & (2)   & (3)                  & (4) & (5) & (6)                & (7) & (8) & (9) \\ \hline

\endhead

\hline
\endfoot
\hline
\caption{
Column 1: galaxy name. Column 2: distance. Different values are assumed in the CO vs. stellar studies for NGC~315 (this paper) and NGC~4697; the surface brightness fluctuation distance is adopted here and all measurements are scaled to this value.
Column 3: black hole mass from stellar-based measurements. Column 4: stellar mass-to-light ratio for the stellar-based measurements (band indicated in square brackets). Column 5: inclination angle assumed in axisymmetric orbit modeling; only \ngc\ is modeled with a triaxial orbit code. Column 6: black hole mass from CO-based measurements. Column 7: stellar mass-to-light ratio for the CO-based measurements (band indicated in square brackets). Column 8: inclination from CO-based measurements. Column 9: references. (a) This work, (b) \citet{boizelleetal2021}, (c) \citet{krajnovicetal2009}, (d) \citet{smithetal2019}, (e) \citet{ruslietal2011}, (f) \citet{barthetal2016}, (g) \citet{watersetal2024}, (h) \citet{boizelleetal2019}, (i) \citet{schulzegebhardt2011}, (j) \citet{davisetal2017}, (k) \citet{ruslietal2013}, (l) \citet{dominiaketal2024a}, (m) \citet{kabasaresetal2022}.\\
$^{\dagger}$ The reported errors for NGC~524 parameters are $3\sigma$ regions; we divide them by $3$ to approximate the $1\sigma$ regions, as quoted for all other galaxies.
}
\label{tab:summary}
\endlastfoot
NGC~315              & 65.9    & $3.0{\pm}0.2$            & $2.49{\pm}0.06$ [J]           & N/A (triaxial)           & $1.96^{+0.30}_{-0.13}$              & $1.86{\pm}0.01$ [J]          & $74.1^\circ{\pm}0.1^\circ$                  & (a,b)   \\[0.5ex]
NGC~524$^{\dagger}$  & 23.3    & $0.83^{+0.09}_{-0.04}$  & $5.8{\pm}0.4$ [I]                      & $20^\circ$ (fixed)  & $0.40^{+0.12}_{-0.07}$             & $5.7{\pm}0.3$ [I]           & $20^\circ$ (fixed)                        & (c,d)   \\[0.5ex]
NGC~1332           & 22.3    & $1.45{\pm}0.2$           & $7.08{\pm}0.39$ [R]                     & $90^\circ$ (fixed)  & $0.664^{+0.065}_{-0.063}$          & $7.83$ [R]            & $85.2^{\circ}$                           & (e,f)  \\[0.5ex]
NGC~3258           & 31.9    & $2.2{\pm}0.2$            & $2.5{\pm}0.1$ [H]                       & $48^\circ$ (fixed)  & $2.249{\pm}0.27$                     & $\cdots$ [H]              & $27.5^\circ {-} 49.3^\circ$                                  & (g,h)  \\[0.5ex]
NGC~4697           & 11.4    & $0.18{\pm}0.05$          & $4.3{\pm}0.3$ [V]                       & $90^\circ$ (fixed)  & $0.13^{+0.003}_{-0.006}$           & $2.14^{+0.04}_{-0.05}$ [i] & $76.1^\circ{}^{+0.5^\circ}_{-0.4^\circ}$ & (i,j)  \\[0.5ex]
NGC~4751           & 26.9    & $1.4{\pm}0.1$            & $12.2^{+0.6}_{-0.7}$ [R]              & $90^\circ$ (fixed)  & $3.43^{+0.16}_{-0.16}$             & $2.68{\pm}0.11$ [H]         & $78.7^\circ{}^{+0.1^\circ}_{-0.1^\circ}$  & (k,l)  \\[0.5ex]
NGC~6861           & 27.3    & $2.0{\pm}0.2$            & $6.1_{-0.1}^{+0.2}$ [I]               & $90^\circ$ (fixed)  & $1{-}3$                           & $2.14{-} 2.52$ [H]           & $72.7^\circ{-} 73.6^\circ$                              & (k,m)  \\
\end{longtable*}
}

\section{Summary}
\label{sec:seven}

We have performed triaxial stellar orbit modeling of the massive elliptical galaxy \ngc\ using photometric data and 
${\sim} 2300$ spatially resolved stellar kinematic measurements in 304 bins covering a radial range of ${\sim}0.3''$ to $30''$ from the MASSIVE survey as constraints.   
After searching over ${\sim}15,000$ galaxy models with an efficient Bayesian scheme, we are able to simultaneously constrain \ngc's \mbh, \ml, dark matter halo mass, and intrinsic shape parameters.

We find \ngc\ to be a triaxial and highly prolate galaxy with a triaxiality parameter $T=0.89 {\pm} 0.02$, hosting a SMBH with $\mbh=(3.0{\pm}0.3){\times} 10^{9} M_\odot$.
At this dynamically inferred mass,
the \ngc\ SMBH is located below  the mean $\mbh{-}\sigma$ and $\mbh{-}M_{\rm bulge}$ scaling relations formed by other local SMBHs and their host galaxies, but it lies within the intrinsic scatter of these relations.  In comparison, the SMBH mass inferred from CO kinematics is $\mbh=(1.96^{+0.30}_{-0.13}) {\times} 10^{9} M_\odot$ (scaled to our distance).

The orbit-based \mbh\ determination reported in this paper adds a measurement to a small but growing sample of galaxies for which the mass of the central SMBH has been measured using more than one dynamical tracers.
Comparing \mbh\ values for a sample of seven galaxies with both CO-based and stellar-orbit based measurements,
we find that the one-to-one relation with an intrinsic scatter term has roughly the same support as a linear relation with free slope, intercept, and intrinsic scatter.
The current data therefore do not indicate statistically significant biases between the masses inferred from the two methods.

At our best estimates of  $\mbh=3.0{\times}10^9 \msun$ and $D=65.9$ Mpc, the \ngc\ SMBH has an angular shadow size of $\theta=2\sqrt{27} G\mbh/c^2 D\approx 4.7\,\mu{\rm as}$. 
Together with its relatively high millimeter flux, \ngc\ is a prime candidate target for future event horizon scale imaging missions (e.g., \citealt{johnsonetal2024}; 
\citealt{zhangetal2024};
\citealt{benzinebetal2024}). A successful measurement of the shadow size would provide another independent estimate of this SMBH's mass. 

\newpage
\acknowledgments

We acknowledge support from NSF-AST-2206219, NSF AST-2206307, and the Heising-Simons Foundation.
This work used the Extreme Science and Engineering Discovery Environment (XSEDE) at the San Diego Supercomputing Center through allocation AST180041, which is supported by NSF grant ACI-1548562.  Portions of this research were conducted with the advanced computing
resources provided by Texas A\&M High Performance Research Computing.
This work is based in part on data obtained at the international Gemini Observatory, a program of NSF’s NOIRLab, which is managed by the Association of Universities for Research in Astronomy (AURA) under a cooperative agreement with the National Science Foundation on behalf of the Gemini partnership: the National Science Foundation (United States), the National Research Council (Canada), Agencia Nacional de Investigación y Desarrollo (Chile), Ministerio de Ciencia, Tecnología e Innovación (Argentina), Ministério da Ciência, Tecnologia, Inovações e Comunicações (Brazil), and Korea Astronomy and Space Science Institute (Republic of Korea).  This work is based in part on observations made with the NASA/ESA Hubble Space Telescope, obtained at the Space Telescope Science Institute, which is operated by the Association of Universities for Research in Astronomy, Inc., under NASA contract NAS5-26555. These observations are associated with program GO-14219.

\software{Astropy \citep{astropycollaborationetal2013, astropycollaborationetal2018},
Dynesty \citep{speagle2020},
Galfit \citep{pengetal2002},
jampy \citep{cappellari2008},
linmix \citep{kelly2007},
Matplotlib \citep{hunter2007},
mgefit \citep{cappellari2002},
NumPy \citep{harrisetal2020}.}

\newpage
\section{Appendix}

\begin{figure}[htp]
\includegraphics[width=\columnwidth]{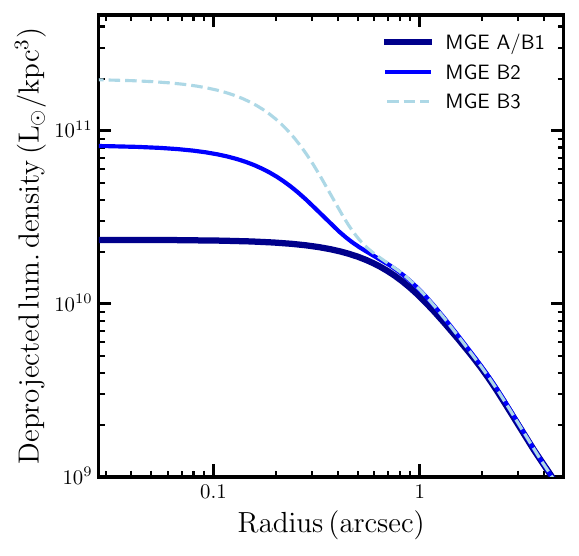}
\caption{Comparison of the deprojected 3D luminosity density for the three MGE models of \ngc\ presented in \citet{boizelleetal2021}. 
The large bumps in the central 3D luminosity densities of model B2 and B3 are an artifact of the 
small widths of the central Gaussian component of these two models,  $\sigma^\prime = 0\farcs178$ and $0\farcs119$, respectively. In comparison, model A has $\sigma^\prime = 0\farcs 580$ for the central component and is well behaved upon deprojection.}

\label{fig:luminosity_density} 
\end{figure}

\bibliography{main}

\end{document}